# BigFCM: Fast, Precise and Scalable FCM on Hadoop

Nasser Ghadiri[1][*], Meysam Ghaffari[2], Mohammad Amin Nikbakht[1]

{nghadiri@cc.iut.ac.ir, ghaffari@cs.fsu.edu, ma.nikbakht@ec.iut.ac.ir}

[1] *Department of Electrical and Computer Engineering, Isfahan University of Technology, Isfahan, Iran*
[2]*Department of Computer Science, Florida State University, Tallahassee, Florida, United states*

**Abstract:** Clustering plays an important role in mining big data both as a modeling technique and a preprocessing step in many data mining process implementations. Fuzzy clustering provides more flexibility than non-fuzzy methods by allowing each data record to belong to more than one cluster to some degree. However, a serious challenge in fuzzy clustering is the lack of scalability. Massive datasets in emerging fields such as geosciences, biology and networking do require parallel and distributed computations with high performance to solve real-world problems. Although some clustering methods are already improved to execute on big data platforms, but their execution time is highly increased for large datasets. In this paper, a scalable Fuzzy C-Means (FCM) clustering named BigFCM is proposed and designed for the Hadoop distributed data platform. Based on the map-reduce programming model, it exploits several mechanisms including an efficient caching design to achieve several orders of magnitude reduction in execution time. Extensive evaluation over multi-gigabyte datasets shows that BigFCM is scalable while it preserves the quality of clustering.

*(*) Corresponding author. Address: Department of Electrical and Computer Engineering, Isfahan University of Technology, Isfahan, Iran. Phone : +98-31-3391-9058, Fax: +98-31-3391-2450, Alternate email: nghadiri@gmail.com*

# 1   Introduction

Billions of computer devices, sensors, biological laboratory experiments and many other data sources generate massive amounts of data. The volume of data is increasing exponentially such that the generated data in each year is equal to all data accumulated in the previous years. There is a growing need for developing efficient and scalable methods to analyze these data is important. Due to the volume and velocity of the big data, using unsupervised methods such as clustering to discover unknown patterns is useful in many modeling and data preprocessing tasks. Clustering is an unsupervised learning method for partitioning the data into a set of groups, such that the records in each group are more similar to each other, and the records in different groups have more discrepancy. Clustering is widely used in data mining to build a standalone model or as a preprocessing method for other modeling and data analysis methods when the dataset is unlabeled. It can be used to discover patterns in vast amounts of data without any need to prior knowledge but without explaining the underlying patterns. Clustering has been widely used in many research domains such as social networks, biomedicine, bioinformatics, engineering, natural language processing and image processing [1]. There are various clustering methods including partitional and hierarchical clustering. In the partitional clustering approach, the algorithm gets the initial number of clusters and starts with initial centers and tries to optimize data clustering based on the given centers such that a predefined objective function is minimized. Two well-known methods in this category are the K-Means clustering that assigns each record to a single cluster, and the Fuzzy C-Means (FCM) clustering that assigns each record to possibly more than one cluster with different fuzzy membership degrees. FCM generally provides better results for the data analyst and gives a more general picture of the dataset [2]. Many improvements have been suggested for the FCM such as methods for selecting better initial seeds compared to the random initialization [3], considering the interaction between clusters to achieve more precise results [4], and using multiple kernel methods [5]. However, when it comes to clustering for big datasets, a serious challenge of the FCM is its scalability. In every iteration of FCM, it requires processing all the data records to compute the initial centers. So it would be very limited for working with big data.

To the best of our knowledge, there are few existing methods for handling the scalability challenge of the FCM using widely-used big data platforms like the Hadoop ecosystem based on the map-reduce programming model. Some methods are based on sampling a few data records while others try to process the whole dataset.

However, most of the existing methods for improving the scalability of FCM have at least one of the following shortcomings:

1- Using the sampling method would speed up the FCM, but it adversely affects the precision of the clustering [6].

2- Executing one map-reduce job per iteration of the FCM, that will require a significant number of jobs to be run with excessive time consumption.

3- Partitioning the dataset and running the core FCM on each partition, and combining the results. This will deteriorate the accuracy of the results because, in the combining phase, the importance of each intermediate result is not considered.

4- Having no idea about decreasing the iterations of the algorithm which makes the algorithm highly time consuming. Some methods just limit the number of iterations. This causes a degradation in the accuracy.

In this paper, to address the aforementioned problems, based on the following contributions. First, we propose a novel approach to analyze the entire collection of records (not just the samples). Thus, unlike many of the existing methods, the final result is based on the complete dataset. Second, in our proposed method, just one map-reduce job works iteratively, and there is no need to execute new jobs. It would provide a substantial improvement in computation time. Third, using a weighted FCM algorithm, the extracted result of processing each part of data has the proper weights showing the importance of each calculated center in that region. Finally using a fast pre-clustering, the number of algorithm iterations is reduced significantly without any adverse effect on the final result. The importance of the number of iterations is important considering the fact that the algorithm must analyze several terabytes and even petabytes of data, and thus, even one iteration may take significant time to execute.

The paper is organized as follows. An overview of existing work is given in Section 2. The proposed BigFCM method is introduced in Section 3. The design of experiments for evaluation of the performance and precision of the method are presented in Section 4 as well as the interpretation of the results. Section 5 concludes the paper and points to some future works.

## 2 Related work

Distance-based clustering is a widely used unsupervised learning approach. K-Means and Fuzzy C-Means (FCM) clustering are two popular methods in this practice. K-Means is a hard clustering that assigns each data record to a single cluster, while FCM is based on fuzzy sets, allowing each data record to be a member of more than one cluster with different degrees of membership [1]. FCM is applied to a variety of domains including bioinformatics [7], image segmentation [8], geology [9] and intrusion detection [10, 11]. An evaluation of K-Mean and FCM shows that FCM performs better in terms of quality. However, K-Means is less time-consuming [12].

Fuzzy clustering of big data has been a challenging task in recent years. Astronomy, cancer diagnosis [13], telecom [14], bioinformatics and internet of things are among the top application areas. A recent application of FCM for network intrusion detection also uses FCM [15] that shows better precision compared to other methods.

To overcome the memory limitation for executing FCM over large datasets, Hore, et al. proposed an algorithm called weighted FCM [16] where the whole dataset is scanned in one pass. Farmstrom, et al. [17] also proposed a single-pass K-Means algorithm that follows a partial data access approach. The weighted FCM was used for applications like image segmentation [16].

Since FCM choose the initial centers randomly, the final result and especially its convergence speed significantly depends on the original center selection. A method proposed to address this problem is based on estimated subsample size to improve the initialization [18]. In the field of clustering large amounts of data, three types of methods have been proposed:

1- Sampling based methods: in these kinds of algorithms a small subset of the dataset is selected, and the clustering is executed on this subgroup [19]. Although this approach is a fast method but the final result is not precise since the whole dataset has not been processed. The precision is an important aspect that must not be ignored in fuzzy clustering [20].

2- Data transformation algorithms: in these methods the structure of the data is altered such that it can be processed more efficiently. These methods are mostly used for graph based structures [21, 22].

3- Single-pass algorithms: in these type of algorithms, the data is divided into subsets and the algorithm loads each chunk and clusters it and then combine the clustering results. The algorithms in this scope are two types:

    a. Incremental clustering methods [23, 24]

    b. Divide and conquer methods [25, 26]

An online weighted improvement of FCM (OWFCM) is proposed in [6] that uses a density-based method to assign the weight to the points. This process is designed to run on a single machine as an online algorithm with no parallel execution. Moreover, it requires a preprocessing to run over the whole dataset to calculate the weights. This density-based preprocessing has a high time complexity of $O(n^2)$.

Ludwig proposed a method for clustering big data using map-reduce model [27]. But the proposed model suffers from executing map and reduce jobs iteratively and hence it has a significant runtime. Even the proposed method is slower than mahout K-Means and fuzzy K-Means. But the mahout K-Means and fuzzy K-Means use the vectors, so they need a preprocessing phase. Although the preprocessing time is much lower than its effect on the overall execution time, so overall the mahout method works better.

However as explained in our methods all of them except sampling based methods, suffer from very long running times. The weakness of sampling based methods is their lower accuracy. Because they execute jobs for each iteration, and furthermore, the primary centers are chosen randomly in these methods. Thus, the convergence process is slow.

## 3  Proposed Method

As discussed in Section 1, the existing FCM methods for the big data suffer from low accuracy and high computational costs. In the proposed method we are going to suggest the BigFCM, a fast, precise and scalable way for big data clustering. In the following after a short introduction on map-reduce programming model and Hadoop framework, the basic FCM and WFCM methods are described briefly and then the BigFCM will be explained.

### 3.1  The Map-Reduce model and the Hadoop framework

One of the widely-used frameworks for working with big data is Hadoop that is proposed to work with the map-reduce programming model. In this frawework, the dataset is initialized, and the algorithm solves the problem in two steps. Initially, the dataset is segmented into predefined subsets in the Hadoop Distributed File System (HDFS). The first function (map) is executed on each partition of the data, and the results are sent to the reduce function. The reduce function aggregates the results, and final results are extracted and written to the HDFS. In some algorithms, it is possible to execute this steps iteratively. Since the Hadoop framework is designed to work on the distributed systems with an enormous amount of data and furthermore tolerate against faults, it is one of the best choices for working on the big data. So the proposed method is designed to have the best efficiency on the Hadoop.

### 3.2  The generic FCM algorithm

The FCM clustering method is an unsupervised distance based clustering algorithm that partitions a dataset of $N$ objects into a set of $C$ clusters such that the records in each cluster are similar to each other, and the records of different clusters are distinct from each other. Unlike traditional non-fuzzy clustering, FCM allows every record to belong to more than one cluster, with varying degrees of membership in each cluster. The FCM algorithm gets a collection of records in the n-dimensional space, and attempts to minimize an objective function (1) to assign the records to the most relevant clusters.

$$Q = \sum_{i=1}^{C} \sum_{k=1}^{N} U_{ik}^{m} ||X_k - V_i||^2 \qquad (1)$$

$V_i$ is the calculated center of the $i^{th}$ cluster in each iteration, and U is a partition matrix that stores the membership degree of each record in each cluster. $m$ is the fuzzification coefficient for controlling the fuzziness of the algorithm.

### 3.3　The Weighted FCM

Instead of the classic FCM, we will use a weighted improvement of it as proposed by [28]. The data objects are treated equally in the basic FCM. The weighted FCM (WFCM) algorithm considers a weight for the objects that shows the importance of each object in the clustering. So the optimization of WFCM is shown as (2).

$$Q = \sum_{i=1}^{C} \sum_{k=1}^{N} w_k U_{ik}^m ||X_k - V_i||^2 \tag{2}$$

$w_k$ is the corresponding weight of each record that shows the influence of it in the optimization process.

### 3.4　The Proposed BigFCM Algorithm

The aim of the proposed method is providing a fast and scalable algorithm for clustering big data with little or no decrease in accuracy of the results, and at the same time, preserving the compatibility with Hadoop platform.

The overall process of the BigFCM method is also depicted in Figure 1. The initial cluster centers are determined by selecting a few random samples from the whole datasets. It will be shown that the sampling method leads to better cluster centers compared to randomly determining the centers as used by many methods. The system components will be described as follows.

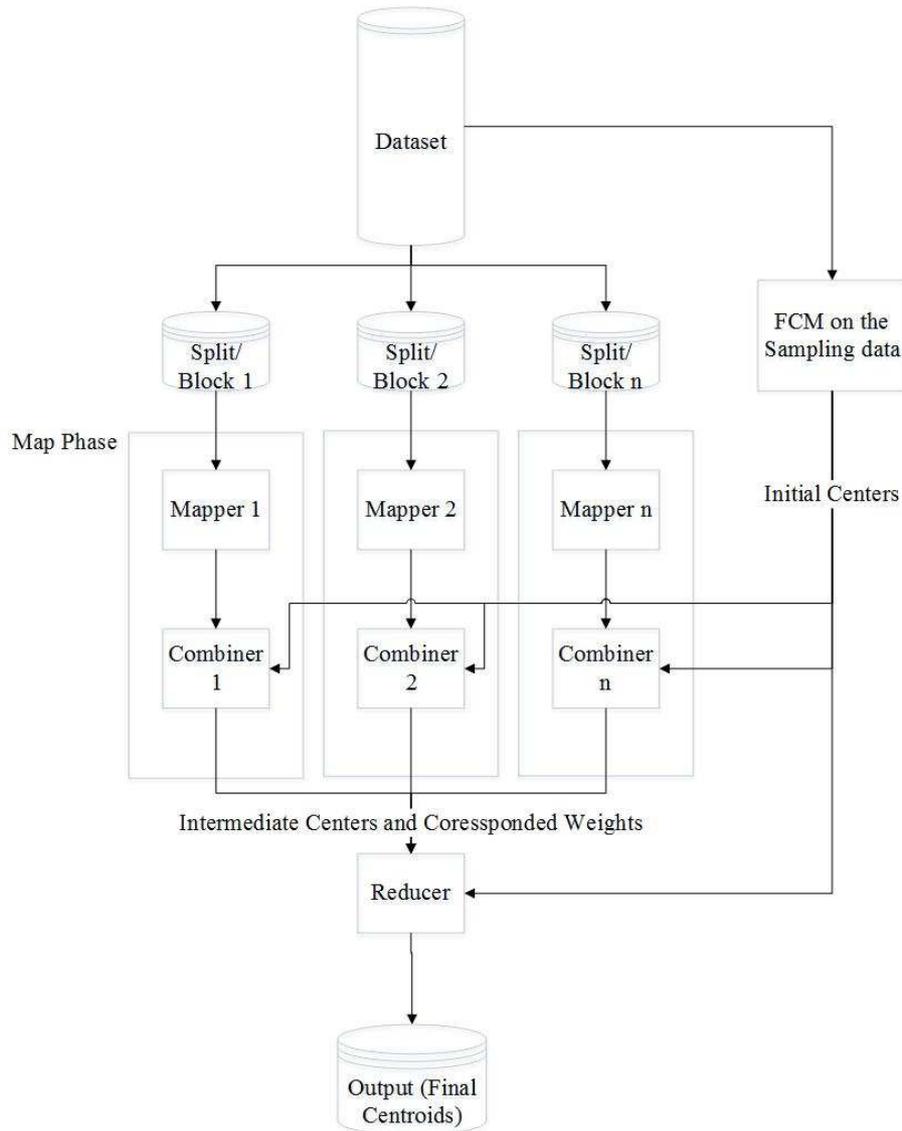

Figure 1. The overall process of BigFCM

In the first step of the BigFCM, a few records are randomly selected from the database in the master node using the Hadoop HDFS options. These records are clustered using basic FCM, and the extracted cluster centers are sent to the Hadoop cache file. Using these initial centers in each iteration will improve the convergence. Each segment of the dataset will be clustered using FCM then the centers and their corresponding weights are sent to the reducer. The weights are used to represent the importance of each center. They show how many records are related to that center, thus improving the precision of the algorithm in subsequent steps. In the last step, the reducer extracts the final centers using WFCM on the output of the previous step. So the proposed BigFCM method is executed as three main phases: initial center extraction, clustering each part of the data using the extracted centers, and finally combining the results using WFCM.

Before explaining the map-reduce steps, a summary of the symbols and notations used in this paper are presented in Table 1.

Table 1. Symbols and notations

| Symbol | Used in which part of the algorithm | Description |
|---|---|---|
| $R_x$ | Driver job | X randomly chosen records for pre-clustering |
| $T_{WFCM}$ | Driver job | WFCMPB running time |
| $T_{FCM}$ | Driver job | FCM running time |
| $R_{subset}$ | Combiner | A subset of data that assigned to the map for clustering |
| $V_{init}$ | Output of the driver job/used in combiner | Extracted initial centers from randomly chosen records that are used as the seeds of the FCM algorithm to speed up convergence |
| $V_{winit}$ | Output of the driver job/used in combiner | Extracted initial centers from randomly chosen records that are used as the seeds of the WFCMPB algorithm to speed up convergence |
| $V_m$ | Output of the combiner used in reducer | The extracted centers of maps and $V_{m\_k}$ is the obtained center of the map k. these centers are sent with corresponding weights to the reduce |
| $V_{final}$ | Output of the WFCMPB/ reducer | The last calculated center of WFCMPB and reducer |
| $C_{Intermediate}$ | Used in the driver job and combiner /WFCMPB | Cluster centers in the intermediate clustering |
| $C$ | Used in the driver /reducer/WFCMB/combiner | Number of desired clusters for the final clustering |
| $W_i$ | Output of the combiner used in the reducer and WFCMPB/used in WFCM | The calculated weight of each center showing the importance of that center based on the corresponding membership value of the records to that center |
| $M$ | Used in the WFCMPB/ driver / reducer / combiner | Fuzzy coefficient. |
| Flag | Output of the driver job used in the combiner | The combiner uses to distinguish between two methods of calculating intermediate centers |

**Distributed Cache file** is an essential feature of the Hadoop accessible by every node. So if the extracted centers in step one are stored in distributed cache file, the Hadoop jobs could use them as first FCM centers, instead of using randomly selected centers. The number of records for the first centers is calculated by Eq. (3) known as Thompson's formula [29]:

$$Smallest\ Sample\ Size = \max_{\mu} z^2 \frac{\left(\frac{1}{\mu}\right)\left(1 - \frac{1}{\mu}\right)}{d^2} \qquad (3)$$

Here $d$ is the maximum absolute difference proportion of each class from the correct proportion and $\mu$ is the number of classes. $Z$ is the upper of $\left(\frac{\alpha}{2\mu}\right) * 100\%$ of the standard normal distribution that shows the acceptable error.

Since the allocation of each class is different, the above formula is not applicable for most datasets, so Parker and Hall proposed Eq. (4):

$$\lambda = \frac{v(\alpha) * c^2}{r^2} \qquad\qquad (4)$$

$r$ is the relative difference between class proportions; $c$ is the number of clusters and $v(\alpha)$ is the value published by Thompson's paper. For example if $\alpha = 0.05$, then $v(\alpha) = 1.27359$. If we have five clusters and the relative difference is 0.10 then based on (4), a total of 3184 records would be required to achieve an error rate below 0.05 [18].

Note that this subsampling has two problems. First, the error rate is theoretical, and the algorithm error will be added, making it even worse. Second, a sound knowledge of every class distribution over the dataset is needed that is not available in many cases.

Based on the above argument, we will use this subsampling method just as an initial estimation facilitator to decrease the execution time of our algorithm, not as a primary component. The FCM algorithm and the sampling are shown in Algorithm 1 are executed for the selected $\lambda$ records based on Eq. (1) optimization function. Their execution times are compared to determine the faster algorithm over the whole dataset.

We can expect that the clustering job will converge faster by using this method. This claim is supported by the fact that the randomly chosen records in the first step are part of the dataset, the extracted centers will provide a good yet coarse view of the dataset partitions in the space. The actual effect of choosing initial centers on the convergence speed will be shown in the experimental results section.

**Map job** is responsible for reading the files from HDFS, eliminating the space or any other user defined separator and making the records ready for processing. This job just reads the data files line by line and sends the preprocessed record to the *combiner*. The output of map job is a combination of (*key*, *value*) pairs. In the proposed fuzzy clustering method, the *key* is used to define the *combiner* in the case of having more than one combiner. If more than one combiner exists, the *key* could be set for each record to assign the corresponding combiner. So the map job sends a combination of the *key* and the record to the combiner. The FCM algorithm will be executed in the combiner job, as an optimized method for this purpose in Hadoop.

**The combiner** is a job usually considered as a part of map phase. This job is executed over the output from the *map* job, and prepares the data for the *reduce* job. In the proposed method, the *combiner* gets the records from the *map* and the initial centroids from the *distributed cache file*. As explained earlier, these centroids are extracted from a subset of data, so they have a good perspective of the real centers. These initial cluster centers will help the algorithm to converge faster. This is crucial in the field of big data since analyzing tera-bytes of data is time consuming, and even a single iteration of fuzzy clustering may take a considerable time. So the computation costs are expected to decrease significantly by using this method. The FCM algorithm is executed by the combiner over the data, and the centroids are calculated based on Eq. (5):

$$numerator_i = (||X_n - V_i||)^{\frac{2}{m-1}} \quad , \quad \forall_i \in \{1, \dots, C\}$$

$$denominator = \sum_{i=1}^{C} \frac{1}{numerator_i}$$ 
(5)

$$U_{i,n} = (numerator_i \times denominator)^{-m} \quad , \quad \forall_i \in \{1, \dots, C\} \text{ // membership term } (U_{in}^{-m})$$

$$V_i = V_i + U_{i,n} \times X_n$$

In Eq. (5), the *numerator$_i$* is computed using Euclidean distance. The *denominator* of the membership function is also calculated based on the primary membership functions. The membership term $U_{i,n}$ and the numerator of centers $V_i$ are then obtained.

This step of the proposed method is shown as Algorithm 1, where *data* is the input records, *C$_{intermediate}$* contains the cluster centers extracted in the previous step or in the driver job. *WP* provides the weights, and *C* shows the number of clusters, *n* is the number of data records, and *M* is the fuzzy coefficient. The reason for using the optimization procedure shown in Algorithm 1 instead of basic FCM is that this method has the time complexity of $O(n.c)$ where *n* is the number of records and *c* is the number of clusters. In contrast, the basic FCM runs with a complexity of $O(n.c^2)$ for such calculations. The Algorithm 1 does not require the membership values, so it is not calculated in this formula, leading to a huge improve in the time complexity.

---

**Algorithm 1: WFCM**

---

**Input**: *data, C $_{intermediate}$, W, C, M*
**Output**: *V$_{final}$, W$_{final}$*

---

**While** max $_{1\leq I \leq C}$ { $||V_{I, new} - V_{I, old}||^2$ } > ε

$\forall_K \in \{1, \dots, \text{number of points}\}$

1. $numerator_i = (||X_k - V_i||)^{\frac{2}{m-1}}$ , $\forall_i \in \{1, \dots, C\}$

   $denominator = \sum_{i=1}^{C} \frac{1}{numerator_i}$

2. $U_{i,k} = (numerator_i \times denominator)^{-m}$ , $\forall_i \in \{1, \dots, C\}$

3. $V_i = V_i + U_{i,n} \times X_n \times W_k$

   $W_i = \sum_{i=1}^{c} U_{i,k} \times W_k$

4. $V_i = \frac{V_i}{W_i}$ , $\forall_i \in \{1, \dots, c\}$

After convergence, the centers and corresponding weights are calculated by Eq. (6):

$$W_{\text{final}} = \sum_{i=1}^{c} U_{i,n}$$

$$V_{final} = \frac{V_i}{W_{final}} \text{ , } \forall_i \in \{1, \dots, c\} \quad // \text{ calculate center (update center)}$$

(6)

The $W_{final}$ weight shows the importance of the centroid which is used in the next step. The output of the combiner is a combination of centroids and corresponding weights. In specific cases that the FCM in the distributed cache file does not converge fast, our experiments show that using the following algorithm in the combiner will be quicker than the Eq. (6). So the based on the execution time of the two FCM versions and Algorithm 1, firstly it determines which algorithm needs to be executed in the combiner. This will depend on the dataset and number of clusters.

**Reduce job** is executed on the output of the maps (*combiners*). As explained earlier in this section, the *reducer* gets a combination of centroids and weights from multiple map processes. The *reducer* then executes a weighted FCM (WFCM) on these centers and extract the desired centers. In the case of having more than one *reducer*, the output of all reducers will be sent to a single reducer that integrates the results by executing WFCM on them to calculate the final centers. The overall process of the WFCMB algorithm is represented in Algorithm 2.

**Algorithm 2: WFCMPB**

**Input**: *data*, $C_{intermediate}$, *C*, *M*
**Output**: $V_{\text{final}}$ , $W_{\text{final}}$

1- Split data to $S_i$ blocks based on sampling formula

2- $V_{final}$ = {}

3- $C_0$ = $C_{intermediate}$

4- $\forall \in \{1, \ldots, S\}$

$\left[ \begin{array}{l} C_i, W_i = \text{FCM } (S_i, C_{i-1}, C, M) \\ V_{final}, W_f = \text{WFCM } (\{V_{final} \cup C_i\}, \{W_f \cup W_i\}, C_{intermediate}, C, M) \end{array} \right.$

After describing the building blocks of our method, now we can explain the overall process of the BigFCM algorithm is depicted in Algorithm 3.

In Algorithm 3, $R_x$ stores the randomly chosen records, and $V_{init}$ contains the extracted initial centers in the driver function that are used as the initial seeds of the FCM in the mappers. $C_{intermediate}$ is the number of intermediate clusters used in the mapper, and $C$ is the number of final desired clusters. We use equal $C_{intermediate}$ and $C$. Here $R_{subset}$ is a subset of all records (a partition) of the whole dataset that is assigned to the map job. $U[i,j]$ is the membership value of each record j in the cluster i. $V_{m\_k}$ is the extracted centers in the mapper $k$, and $W_i$ is the corresponding weight of each center within these centers, and $V_{final}$ is the final obtained centers.

In Algorithm 3, the driver job begins by selecting a small subset of records in line 1, In line 2 and 4, the centers of clustering these records are computed using FCM per block and basic clustering respectively with $C_{intermediate}$ clusters. Their execution time is compared in line 6 to choose the better method for the current dataset, and the extracted centers are stored in the Hadoop distributed cache file. In lines 7-9, the mapper reads each row of the assigned partition and sends the records with the standard format of the Hadoop (*key*, *value*) to the combiner.

---

**Algorithm 3: The BigFCM Algorithm**

---

**Input**: *data, $C_{intermediate}$, C, M*

**Output**: the final extracted centers ($V_{final}$)

---

**Driver job**:

1- Choose $R_x$ random numbers from the HDFS based on sampling size formula

2- $V_{winit}$ = WFCMPB($R_x$, $C_{intermediate}$, C, M)

3- Calculate $T_f$

4- $V_{init}$ = FCM($R_x$, $C_{Intermediate}$, M)

5- Calculate $T_s$

6-  If ( $T_f$ - $T_s$ > 0)
   ⌐ Flag =1
   └ Send $V_{init}$ to the Cache file
   Else
   ⌐ Flag =0
   └ Send $V_{winit}$ to the Cache file

**Mapper:**

7-  Read each record from the file
8-  Eliminate spaces, comma
9-  Send each record for the combiner (*Key*, *Record*)

**Combiner**:

10- If ( Flag equals  1 )
   $V_{m\_k}, W_k$ = FCM($R_{subset}, V_{init}, C, M$)
   Else
   $V_{m\_k}, W_k$ = WFCMPB($R_{subset}, V_{winit}, C, M$)
11- Out: (*key*,{ $V_{m\_k}, W_k$})

**Reducer**:

*12*- Get all $V_i, W_i$
13- $V_{final}, W_f$=WFCM ({$V_1$ U ...U $V_k$}, {$W_1$ U...U $W_k$}, $V_1, C, M$)
14- Out:(*key*, $V_{final}$ )

---

The combiner executes the FCM on the records using the calculated initial centers obtained quickly from the distributed cache file. Then it extracts the centers of these records ($V_{m\_k}$) in line ١٠ and 11, followed by calculating the corresponding weights for each center. The weight of each center is equal to the sum of membership values of all records of the current subset to that center. In line 11, the combiner sends the results for the reducer. In line 12, the reducer gets the extracted centers and their corresponding weights from all mappers, and in line 13 it executes a WFCM on these records using the number of clusters *C*. In line 14, the output will be written to the HDFS. If there are a large number of intermediate centers ($V_m$), it would be possible to execute multiple reduce jobs (like combiner but instead of basic FCM, the WFCM must be used) and then integrate the results like the explained reduce job.

## 3.5   Evaluation metrics

The comparison could be made based on two objectives: (1) measuring the execution time of the each algorithm over the datasets of different sizes, and (2) measuring the precision of the results in each situation. The metrics that we used for this purpose were relative speedup, execution time, and confusion matrix [18]. The relative speedup shows the relative speed of two algorithms which is directly related to the execution time of the algorithms. The speedup and running time criteria are essential to measure the scalability of the system. However, since the accuracy may decrease as a side-effect of performance improvement attempts, the confusion matrix will be used to evaluate the effect of the proposed algorithm on the accuracy. Moreover, the silhouette width criterion helps us to evaluate the quality of clustering algorithms by showing how the clusters are cohesive inside and well-separated from other clusters [30].

# 4   Experimental results

In this section, the proposed BigFCM is analyzed using multiple datasets including classic small-size datasets for necessary testing, and a few large datasets for evaluating the scalability of the method. The evaluation criteria are also described and applied in the experiments.

## 4.1   Experiment design

The proposed algorithm evaluated on a Hadoop cluster of Intel Core i5 computers with 4GB of RAM. The Pima Indian Diabetes and Iris datasets from UCI repository of machine learning[1] were used to evaluate the operation of the algorithm with different parameter configurations. Furthermore, the experiments were also conducted on the public KDD99 network intrusion detection dataset[2]. The KDD99 dataset was normalized and convert categorical features into numerical. For evaluating the scalability of BigFCM, two multi-gigabyte datasets were selected from the same repository. The first is SUSY[3] dataset with 4 million records and 18 features for each record, and the second is HIGGS[4] dataset with 11 million records and 28 attributes for each record. The effect of changing the parameter values in the proposed method is evaluated, followed by comparing BigFCM with the state of the art methods, Apache Mahout K-Means and Fuzzy K-Means. The reason for this comparison is that these two approaches are the most developed techniques in this field. In the following subsections, first, the effect of using subsampling clustering in the driver job of the BigFCM is examined. Then the comparison of the BigFCM and other methods would be investigated.

## 4.2   Results and discussion

The following experiments were conducted based on the aforementioned test design and evaluation criteria.

### 4.2.1   The effect of changing epsilon in the driver

As described in Section 3, the proposed BigFCM runs a primary clustering over the sampled records in the driver to extract initial centers for the mappers (*combiners*) and to find the better fuzzy clustering algorithm for the current dataset. As already described, the FCM algorithm starts with random initial centers and tries to improve the centers in an iterative process. So it can be quickly followed that if the initial centers are placed closer to the optimum values, the algorithm will converge faster. More precise initial centers will help the *combiner* component to execute faster, leading to lower execution time in the proposed method compared to other methods. Therefore, decreasing the execution time of this element will put a significant effect on the overall process. In the *combiner*, the effect of using smaller target *epsilon* (the precision value that determines the minimum change in cluster center as a stop condition) in the driver will be analyzed. Having smaller driver will lead to extracting more precise initial centers based on the sampled records. Table 2 shows the effect of having smaller *epsilon* on the overall execution time of the proposed method on the SUSSY dataset. Note that when we use a smaller *epsilon* in the driver, the execution time of the driver will be increased. However, since the driver job is executed on a small portion of the dataset, and this more precise value will have a significant effect on the future steps, the overall execution time will decrease significantly.

---

[1] http://archive.ics.uci.edu/ml/
[2] http://kdd.ics.uci.edu/databases/kddcup99/kddcup99.html
[3] http://archive.ics.uci.edu/ml/datasets/SUSY
[4] http://archive.ics.uci.edu/ml/datasets/HIGGS

Table 2. Effect of changing epsilon in the driver on the total execution time of the proposed method

| Dataset | Method | Parameters | Random Seed | Initial Epsilon= 5.0E-6 | Initial Epsilon= 5.0E-8 | Initial Epsilon= 5.0E-10 | Initial Epsilon= 5.0E-11 |
|---------|--------|-----------|-------------|------------|------------|------------|------------|
| **SUSY** | BigFCM | iterations =1000, | 5432 | 3038 | 2051 | 918 | 882 |
| | | Reducer Epsilon | Seconds | Seconds | Seconds | Seconds | Seconds |
| | | = 5.0E-11, M=2 | (90 mins) | (50 mins) | (34 mins) | (15 mins) | (14 mins) |
| | | Centroid = 10 | | | | | |

As shown in Table 2, if the initial center calculation is not used and the traditional FCM is executed in the combiners, the execution time will be about 90 minutes. However, in the case of using the sampling and initial center extraction in the driver, based on the epsilon execution time will be decreased. The significant difference between these two values shows that using initial center extraction with higher precision could lead to more than 6 times speedup.

### 4.2.2   *Comparing execution time of BigFCM with other methods*

In this subsection, the execution time of the BigFCM with different methods will be compared. This comparison is based on the target precision (Epsilon), the size of the data and the number of clusters. Table 3 shows the execution time of Mahout K-Means, Fuzzy K-Means and the BigFCM over SUSY and HIGGs datasets.

Table 3. Execution time of BigFCM, Mahout K-Means and Mahout Fuzzy K-Means over the SUSY dataset

| Dataset | Method | Parameters | Epsilon = 5.0E-7 (s) | Epsilon = 5.0E-5 (s) | Epsilon = 5.0E-3 (s) | Epsilon = 5.0E-2 (s) |
|---------|--------|-----------|------|------|------|------|
| **SUSY** | Mahout FKM | C = 2 , iterations =1000 | 141887 | 4308 | 3000 | 930 |
| | Mahout KM | C =2 , M =2, iterations = 1000 | 2328 | 1680 | 1025 | 710 |
| | BigFCM | C =2 , M =2 , iterations = 1000 | 435 | 436 | 432 | 430 |
| **HIGGS** | Mahout FKM | C =2 , M =2, iterations = 1000 | 6120 | 3996 | 3287 | 1848 |
| | Mahout KM | C = 2 , iterations =1000 | 4430 | 4446 | 4434 | 2568 |
| | BigFCM | C =2 , M =2 , iterations = 1000 | 480 | 480 | 475 | 473 |

The results in Table 3 show that BigFCM runs much faster than other algorithms. The important point is that as shown in Fig. 2 using lower epsilon has a negligible effect on BigFCM. This is due to the initial center extraction step of the proposed method. So by using the BigFCM, it will not be needed to sacrifices the precision, in order to achieve lower execution time.

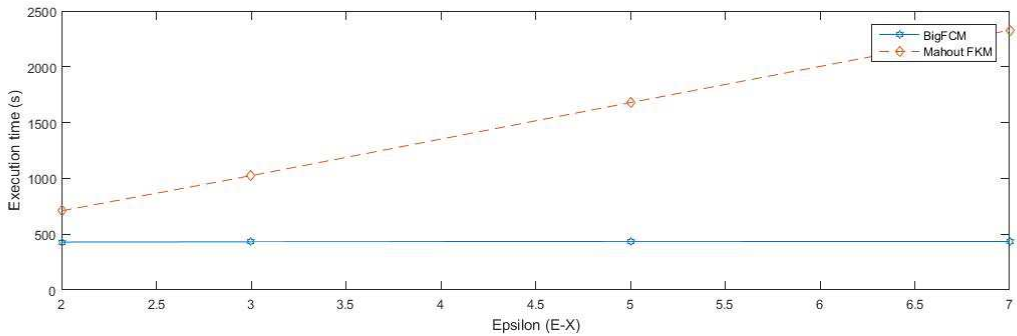

Figure 2. Effect of changing Epsilon on BigFCM and Maout FKM on SUSY dataset

This figure shows that unlike Mahout FKM, in order to have a more precise algorithm, the execution time of the BigFCM does not change significantly.

In Table *4*, the execution times of the methods are compared based on having different size of the data. The data shows a number of the records and file size shows the analyzed data size. The results in this table indicate that the proposed BigFCM method provides the speedup values of 287 and 493 times the Mahout K-Means and Fuzzy K-Means with the same parameters, respectively. These results also depicted in Fig. 3. It can be observed that BigFCM performs 2GB of data clustering about 100 times faster than clustering 50MB of data using Mahout K-Means or Mahout FKM.

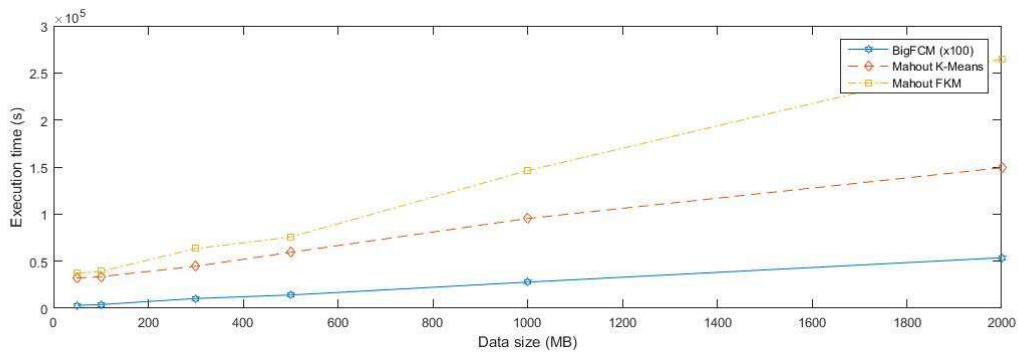

Figure 3. The execution time of the BigFCM (x100) , Mahout K-Means and Mahout FKM for different data sizes. The execution times of the BigFCM are multiplied by 100 for better illustration.

Table 4. Execution time of the BigFCM, Mahout K-Means and Mahout Fuzzy K-Means for different data sizes

| Size of Data | File size in Bytes | BigFCM<br>C =6 , Epsilon = 5.0E-11 ,<br>M =2,<br>Iterations = 1000<br>(s) | Mahout KM<br>C =6 , Epsilon = 5.0E-11 ,<br>Iterations = 1000<br>(s) | Mahout FKM<br>C =6 , Epsilon = 5.0E-11 ,<br>M =2 ,<br>Iterations = 1000<br>(s) |
|---|---|---|---|---|
| 20K | 10485760 | 18 | 31468 | 31620 |
| 40K | 20971520 | 30 | 32268 | 32280 |
| 60K | 31457280 | 30 | 32418 | 33394 |
| 80K | 41943040 | 34 | 31709 | 35600 |
| 100K | 52428800 | 31 | 32160 | 37080 |
| 120K | 62914560 | 39 | 32454 | 35820 |
| 140K | 73400320 | 41 | 32301 | 36480 |
| 160K | 83886080 | 39 | 32731 | 37140 |
| 180K | 94371840 | 42 | 33414 | 38080 |
| 200K | 104857600 | 40 | 33589 | 39286 |
| 400K | 209715200 | 73 | 36601 | 47053 |

| 600K | 314572800 | 104 | 44718 | 63480 |
|------|-----------|-----|-------|-------|
| 800K | 419430400 | 120 | 47727 | 68920 |
| 1M | 524288000 | 141 | 59418 | 75652 |
| 1.2M | 629145600 | 166 | 60278 | 89152 |
| 1.4M | 734003200 | 207 | 70518 | 103842 |
| 1.6M | 838860800 | 233 | 73744 | 111378 |
| 1.8M | 943718400 | 253 | 85672 | 129994 |
| 2M | 1048576000 | 278 | 95413 | 146229 |
| 3M | 1572864000 | 428 | 121854 | 235796 |
| 4M | 2097152000 | 537 (9 mins) | 149316 (1d 17h 28m) | 264974 (3 days) |

Another important factor in the clustering execution time is the number of the clusters. Table 5 shows the execution time of the BigFCM method for a different number of clusters. The running time of K-Means and Fuzzy K-Means for their best cases were more than 41 hours and 72 hours respectively.

As this table shows the effect of increasing the number of clusters on the proposed method is linear. The reason is that in the combiner part, instead of executing the traditional FCM, a modified version of the FCM has been used.

Table 5. Execution time of the BigFCM for different number of clusters

| Dataset | Method | Parameters | Execution Time (s) | | | |
|---------|--------|------------|---------------------|---|---|---|
| | | | Centroid = 6 | Centroid = 10 | Centroid = 15 | Centroid = 50 |
| **HIGGS** | BigFCM | Iterations =1000, Epsilon = 5.0E-11, M=2 | 537 | 2057 | 2970 | 4332 |

Since the mahout K-Means and mahout fuzzy K-Means do not respond in the meantime on our cluster, we could not specify their precise execution time, but in their execution time is much more than the reported time such that the jobs fail during the long-time process hazards.

Furthermore, Table 6 shows a general comparison of the BigFCM and Fuzzy K-Means on various datasets.

Table 6. Execution time of the BigFCM, Mahout K-Means for different datasets

| Dataset | Parameters | Mahout FKM (s) | BigFCM (s) |
|---|---|---|---|
| SUSY | Iterations =1000, Reducer Epsilon = 5.0E-7, M=2, Centroid = 2 | 2328 | 435 |
| HIGGS | Iterations =1000,  Epsilon = 5.0E-7, M=2, Centroid = 2 | 6120 | 480 |
| Pima Indian Diabetes | Iterations =1000, Reducer Epsilon = 5.0E-2, M=1.2, Centroid = 2 | 222 | 5 |
| Iris | Iterations =1000, Reducer Epsilon = 5.0E-2, M=1.2, Centroid = 3 | 66 | 3 |
| KDDCUP 99(10%) | Iterations =1000, Reducer Epsilon = 5.0E-7, M=1.2, Centroid = 23 | 2100 | 300 |

As this table shows, the BigFCM works 5.35 to 44 times (18.22 on average) faster than the Mahout Fuzzy K-Means.

### 4.2.3   Comparing the precision of the results

In this set of experiments, the result of the methods is compared using two metrics, confusion matrix (Table 7) and Silhouette Width (Table 8).

Note that the weak values of the silhouette width metric for the Mahout K-Means are due to the rounding made to enable a faster execution. However, BigFCM focuses on the execution time, while preserving the accuracy of the algorithm.

One can observe in Table 7 and Table 8 that the precision of the results of the BigFCM is better than the Mahout fuzzy K-Means. So the proposed method works faster and more precise than the existing methods.

Table 7. The precision of the results of the algorithms based on confusion matrix

| Dataset | Parameters | Mahout FKM | BigFCM |
|---|---|---|---|
| SUSY | Iterations =1000, Reducer Epsilon = 5.0E-7, M=2, Centroid = 2 | 50.0 % | 50.0 % |
| HIGGS | Iterations =1000,  Epsilon = 5.0E-7, M=2, Centroid = 2 | 50.0 % | 50.0 % |
| Pima Indian Diabetes | Iterations =1000, Reducer Epsilon = 5.0E-2, M=1.2, Centroid = 2 | 65.7 % | 66.1 % |
| Iris | Iterations =1000, Reducer Epsilon = 5.0E-2, M=1.2, Centroid = 3 | 89.1 % | 92.0 % |
| KDDCUP99(10%) | Iterations =1000, Reducer Epsilon = 5.0E-7, M=1.2, Centroid = 23 | 78.0 % | 82.0 % |

Table 8. The precision of the results of the algorithms based on silhouette width

| Dataset | Method | Parameters | 1k | 2k | 3k | 4k |
|---|---|---|---|---|---|---|
| HIGGS | Mahout FKM | Iterations =1000, Epsilon = 5.0E-11, M=2 | 0.0 | 0.0 | 0.0 | 0.0 |
|  | BigFCM | Iterations =1000, Epsilon = 5.0E-11, M=2 | 0.0629 | 0.0637 | 0.0635 | 0.0623 |

# 5  Conclusions

In this paper, a novel clustering method for big data clustering was suggested based on the FCM. The proposed BigFCM method was designed for the Hadoop platform with an optimized map and reduce functions. It was extensively tested and analyzed and compared with the state of the art methods over multiple datasets. The experimental results show that BigFCM executes on the average 18.22 times faster than Mahout Fuzzy K-Means having a mediocre precision. The algorithm could run even more quickly by defining a lower epsilon value. This will enable the proposed BigFCM method to execute up to 493 times faster compared to the Mahout K-Means and Mahout Fuzzy K-Means. Moreover, the BigFCM provides high quality clusters, and the results are more precise than similar methods, making it a more efficient algorithm while preserving the quality of clustering. Future work may include adapting the algorithm to different domains and tuning the required parameters or exploiting domain-specific distance measures.